\documentclass[amsmath]{PoS}

\newcommand{\veps}{{\vec\epsilon}}
\newcommand{\vepsprime}{{\vec\epsilon\, '}}
\newcommand{\vsigma}{{\vec\sigma}}

\def\L{{\Lambda}}

\def\Dslash{D\hskip-0.65em /}

\def\CPT{{$\chi$PT}}

\def\eqref#1{{(\ref{#1})}}

\def\L{{\Lambda}}

\title{Electromagnetic and spin polarisabilities in lattice QCD}
                                                                                                                     
\ShortTitle{Electromagnetic and spin polarisabilities in lattice QCD}
                                                                                                                     
\author{\speaker{W.~Detmold} and A.~Walker-Loud\\
  Department of Physics, University of Washington, Box 351560,
  Seattle, WA 98195, U.S.A.\\
\email{wdetmold@phys.washington.edu}}
\author{B.~C.~Tiburzi\\
  Department of Physics, Duke University, PO Box 90305, Durham, NC
  27708, U.S.A.}

\abstract{We discuss the extraction of the electromagnetic and spin
  polarisabilities of nucleons from lattice QCD \cite{Detmold:2006vu}.
  We show that the external field method can be used to measure all
  the electromagnetic and spin polarisabilities including those of
  charged particles.  We then turn to the extrapolations required to
  connect such calculations to experiment in the context of chiral
  perturbation theory, finding a strong dependence on the lattice
  volume and quark masses.}

\FullConference{XXIVth International Symposium on Lattice Field Theory\\
                July 23-28, 2006\\
                Tucson, Arizona, USA}

\begin{document}

\section{Introduction}

Compton scattering at low energies is an invaluable tool with which to
study the electromagnetic structure of hadrons. 
The real Compton scattering amplitude describing the elastic
scattering of a photon on a spin-half target such as the proton or
neutron can be parameterised as
\begin{eqnarray}
T_{\gamma N}&=&  A_1\, \vepsprime\cdot\veps
            +A_2\, \vepsprime\cdot \hat{k} \, \veps\cdot \hat{k'}
             +i\,A_3\, \vsigma\cdot (\vepsprime\times\veps)
             +i\,A_4\, \vsigma\cdot
             (\hat{k'}\times\hat{k})\, \vepsprime\cdot\veps 
     \nonumber \\
 & & 
     +i\,A_5\, \vsigma\cdot \left[(\vepsprime\times\hat{k})\,
     \veps\cdot\hat{k'} 
            -(\veps\times\hat{k'})\, \vepsprime\cdot\hat{k}\right]
     +i\,A_6\, \vsigma\cdot \left[(\vepsprime\times\hat{k'})\,
     \veps\cdot\hat{k'} 
                        -(\veps\times\hat{k})\,
                        \vepsprime\cdot\hat{k}\right],
\label{eq:Ti}
\end{eqnarray}
where we work in the Breit frame of the system and the incoming and
outgoing photons have momenta $k=(\omega,\vec{k}=\omega\,\hat{k})$ and
$k^\prime=(\omega,\vec{k}^\prime=\omega\,\hat{k}^\prime)$, and
polarisation vectors $\epsilon$ and $\epsilon^\prime$, respectively.
The $A_i=A_i(\omega,\theta)$ are scalar functions of the photon energy
and scattering angle, $\cos\theta=\hat{k}\cdot\hat{k}^\prime$, and can
be separated into two pieces.  The first set, the Born terms, describe
the interaction of the photon with a point-like target with mass,
$M_N$, charge, $e\, Z$ (where $e>0$), and magnetic moment, $\mu$. The
remaining parts of the amplitude describe the structural response of
the target and expanding the amplitude for small photon energies
relative to the target mass and keeping terms to ${\cal O}(\omega^3)$
one can describe the target structure in terms of the electric,
magnetic and four spin polarisabilities \cite{Ragusa:1993rm},
$\alpha$, $\beta$, and $\gamma_{1\mbox{--}4}$, respectively.

Lattice techniques provide a method to investigate the
non-perturbative structure of hadrons directly from QCD.  In
particular, the various hadron polarisabilities can be computed.
Direct calculations of the required hadronic current-current
correlators are difficult on the lattice and so far have not been
attempted. However progress has been made
\cite{Fiebig:1988en,Christensen:2004ca,Lee:2005dq} in extracting the
electric and magnetic polarisabilities by studying the quadratic shift
in the hadron mass that is induced in quenched lattice calculations in
constant background electric or magnetic fields. These studies have
investigated the electric polarisabilities of various neutral hadrons
(in particular, the uncharged vector mesons and uncharged octet and
decuplet baryons), and the magnetic polarisabilities of the baryon
octet and decuplet, as well as those of the non-singlet pseudo-scalar
and vector mesons.  As we shall discuss below, generalisations of
these methods using non-constant fields allow the extraction of the
spin polarisabilities from spin-dependent correlation functions and
also allow the electric polarisabilities to be determined for charged
hadrons. More generally, higher-order polarisabilities and generalised
polarisabilities are accessible using this technique.

As with all current lattice results, these calculations have a number
of limitations and so are not physical predictions that can be
directly compared to experiment. For the foreseeable future, lattice
QCD calculations will necessarily use quark masses that are larger
than those in nature because of limitations in the available
algorithms and computational power. Additionally, the volumes and
lattice-spacings used in these calculations will always be finite and
non-vanishing, respectively.  For sufficiently small masses and large
volumes, the effects of these approximations can be investigated
systematically using the effective field theory of the low energy
dynamics of QCD, chiral perturbation theory ($\chi$PT) \cite{CHPT}.
In this paper we present results for the nucleon electromagnetic and
spin polarisabilities at next-to-leading order (NLO) in the chiral
expansion.  We do so to discuss the infrared effects of the quark
masses and finite volume in two-flavour QCD and its quenched and
partially-quenched analogues (QQCD and PQQCD).  The polarisabilities
are particularly interesting in this regard since they
are very sensitive to infrared physics and their mass and volume
dependence is considerably stronger than that expected for hadron
masses and magnetic moments.

%
%
%
%
%
%
%       Lattice Compton Scattering
%
%
%
%
%
%
\section{Polarisabilities on the lattice}
\label{sec:Compton}

The use of external fields in lattice QCD has a long history. The
pioneering calculations of Refs.~\cite{Fucito:1982ff,
  Martinelli:1982cb,Bernard:1982yu,EDM,Shintani:2005du}
attempted to measure the nucleon axial couplings, magnetic moments and
electric dipole moments by measuring the linear shift in the hadron
energy as a function of an applied external weak or electromagnetic
field.  As discussed in the Introduction, various groups
\cite{Fiebig:1988en,Christensen:2004ca,Lee:2005dq} have also used this
approach to extract electric and magnetic polarisabilities in quenched
QCD by measuring a quadratic shift in the hadron energy in external
electric and magnetic fields. The method is not limited to electroweak
external fields and can be used to extract many matrix elements such
as those that determine the moments of parton distributions and the
total quark contribution to the spin of the proton
\cite{Detmold:2004kw}. Here we focus on the electromagnetic case.

The Euclidean space ($x_4 \equiv \tau$) effective action, $S_{\rm
  eff}[A]=\int d^3 x\,d\tau\, {\cal L}_{\rm eff}(\vec{x},\tau;A)$,
describing the gauge and parity invariant interactions of a
non-relativistic spin-half hadron of mass $M$ and charge $q$ with a
classical U(1) gauge field, $A^\mu(\vec{x},\tau)$, is formed from the
Lagrangian
\begin{eqnarray}
  \label{eq:eff_L}
  {\cal L}_{\rm eff}(\vec{x},\tau;A) &=& \Psi^\dagger(\vec{x},\tau)
  \Bigg[\left(\frac{\partial}{\partial \tau}+i\,q\,A_4\right)
+\frac{(-i \vec\nabla-\,q\,\vec{A})^2}{2M}  - \mu\, \vec{\sigma}\cdot\vec{H}
  +{2\pi}\left(\alpha\, \vec{E}^2 -\beta\, \vec{H}^2\right) 
\\
  &&  \hspace*{-1.8cm}
    -2\pi i \left( -\gamma_{E_1E_1} \vec{\sigma}\cdot\vec{E}\times\dot{\vec{E}}
    + \gamma_{M_1M_1} \vec{\sigma}\cdot\vec{H}\times\dot{\vec{H}}
    + \gamma_{M_1E_2} \sigma^i E^{ij}H^j
    + \gamma_{E_1M_2} \sigma^i H^{ij}E^j
\right)\Bigg] \Psi(\vec{x},\tau)  +\ldots \,,
\nonumber 
\end{eqnarray}
where $\vec{E}=- \frac{\partial}{\partial \tau} \vec{A}(\vec{x},\tau)
-\vec{\nabla} A_4(\vec{x},\tau)$ and $\vec{H}= \vec{\nabla}\times
\vec{A}(\vec{x},\tau)$ are the corresponding electric and magnetic
fields, $\dot{X} = \frac{\partial}{\partial \tau} X$ denotes the
Euclidean time derivative, $X^{ij}=\frac{1}{2}(\partial^i
X^j+\partial^j X^i)$, and the ellipsis denotes terms involving higher
dimensional operators.  The constants that appear in
Eq.~(\ref{eq:eff_L}) are the magnetic moment and electromagnetic and
multipole polarisabilities: $\gamma_{E_1E_1}=-(\gamma_1+\gamma_3)$,
$\gamma_{M_1M_1}=\gamma_4$, $\gamma_{E_1M_2}=\gamma_3$ and
$\gamma_{M_1E_2}=\gamma_2+\gamma_4$.  The Schr\"odinger equation
corresponding to Eq.~(\ref{eq:eff_L}) determines the energy of the
particle in an external U(1) field in terms of the charge, magnetic
moment, and polarisabilities.

Lattice calculations of the energy of a hadron in an external U(1)
field are straight-forward. One measures the behaviour of the usual
two-point correlator on an ensemble of gauge configurations generated
in the presence of the external field. This changes the Boltzmann
weight used in selecting the field configurations from
$\det\left[\Dslash+m\right]\exp{\left[-S_g\right]}$ to
$\det\left[\Dslash
  +i\,\hat{Q}\,\slash\!\!\!\!A+m\right]\exp{\left[-S_g\right]}$, where
$\Dslash\,$ is the SU(3) gauge covariant derivative, $\hat{Q}$ is the
quark electromagnetic charge operator, and $S_g$ is the usual SU(3)
gauge action.  Since calculations are required at a number of
different values of the field strength in order to correctly identify
shifts in energy from the external field, this is a relatively
demanding computational task.
The exploratory studies of
Refs.~\cite{Fiebig:1988en,Christensen:2004ca,Lee:2005dq} used quenched
QCD in which the gluon configurations do not feel the presence of the
U(1) field as the quark determinant is absent. In this case, the
external field can be applied after the gauge configurations had been
generated and is simply implemented by multiplying the SU(3) gauge
links of each configuration by link variables corresponding to the
fixed external field: $\{U^{\mu}_{\alpha}(x)\}\longrightarrow
\{U^{\mu} _{\alpha}(x) \exp[i\, e\,a\, A^\mu]\}$, where $a$ is the
lattice spacing. These studies are interesting in that they provide a
proof of the method, but the values of the polarisabilities extracted
have no connection to those measured in experiment.

It is clear from Eq.~(\ref{eq:eff_L}) that all six polarisabilities
can be extracted using suitable space and time varying background
fields if the shift of the hadron energy at second order in the
strength of the field can be determined.  In order to determine the
polarisabilities, we consider lattice calculations of the two-point
correlation function
\begin{eqnarray}
  \label{eq:correlator}
  C_{s s^\prime}(\vec{p},\tau;A)=\int d^3x\, e^{i\vec{p}\cdot\vec{x}}
\langle 0| \chi_s(\vec{x},\tau)\chi^\dagger_{s^\prime}(0,0) |0\rangle_A\,,
\end{eqnarray}
where $\chi_s(\vec{x},\tau)$ is an interpolating field with the quantum
numbers of the hadron under consideration (we will focus on the
nucleons) with $z$ component of spin, $s$, and the correlator is
evaluated on the ensemble of gauge configurations generated with the
external field, $A^\mu$.

For weak external fields (such that higher order terms in
Eq.~(\ref{eq:eff_L}) can be safely neglected), the small $\vec{p}$ and
large $\tau$ dependence of this QCD correlation function is reproduced
by the equivalent correlator calculated in the effective theory
corresponding to the Lagrangian, Eq.~(\ref{eq:eff_L}). That is
\begin{eqnarray}
  \label{eq:eft_correlator}
  C_{s s^\prime}(\vec{p},\tau;A)&=&\int d^3x\, e^{i\vec{p}\cdot\vec{x}}
\frac{1}{{\cal Z}_{\rm eff}[A]}\int {\cal D}\Psi^\dagger {\cal D}\Psi
\,\Psi_s(\vec{x},\tau)\Psi^\dagger_{s^\prime}(0,0)
\exp\left(-S_{\rm
    eff}[A]\right)\,,
\end{eqnarray}
where ${\cal Z}_{\rm eff}[A]=\int {\cal D}\Psi^\dagger {\cal D}\Psi
\exp\left({-{S}_{\rm eff}[A]}\right)$.  Since the right-hand side of
Eq.~(\ref{eq:eft_correlator}) is completely determined in terms of the
charge, magnetic moment and polarisabilities that we seek to extract,
fitting lattice calculations of $C_{s s^\prime}(\vec{p},\tau;A)$ in a
given external field to the effective field theory expression will
enable us to determine the appropriate polarisabilities.  In the above
equation we have assumed that the ground state hadron dominates the
correlator at the relevant times. For weak fields this will be the
case. However one can consider additional terms in the effective
Lagrangian that describe the low excitations of the hadron spectrum.

In many simple cases such as constant or plane-wave external fields,
the EFT version of $C_{s s^\prime}(\vec{p},\tau;A)$ can be determined
analytically in the infinite volume, continuum limit
\cite{Schwinger:1951nm}. However at finite lattice spacing and at
finite volume, calculating $C_{s s^\prime}(\vec{p},\tau;A)$ in the EFT
becomes more complicated. In order to determine the EFT correlator, we
must invert the matrix ${\cal K}$ defined by
\begin{eqnarray}
  \label{eq:S_latt}
  S_{\rm
  latt}[A]=\sum_{\vec{x},\tau_x}\sum_{\vec{y},\tau_y}\sum_{s,s^\prime}
\Psi^\dagger_s(\vec{x},\tau_x) 
{\cal K}_{ss^\prime}[\vec{x},\tau_x,\vec{y},\tau_y;A] \Psi_s(\vec{y},\tau_y)\,,
\end{eqnarray}
where $S_{\rm latt}[A]$ is a discretisation of the EFT action in which
derivatives are replaced by finite differences. For the most general
space-time varying external field, this must be inverted numerically;
given a set of lattice results for the correlator,
Eq.~(\ref{eq:eft_correlator}) is repeatedly evaluated for varying
values of the polarisabilities until a good description of the lattice
data is obtained.  If the external fields are weak,
$|A^\mu(\vec{x},\tau)|^2\ll\L_{\rm QCD}^2$ for all $\vec{x}$ and
$\tau$, a perturbative expansion of ${\cal K}^{-1}$ in powers of the
field can also be used.

To extract all six polarisabilities using such an analysis, we need to
consider a number of different fields; lattice calculations of the
correlators in Eq.~(\ref{eq:correlator}) (also for different spin
configurations) using the exemplar fields given in
Ref.~\cite{Detmold:2006vu} for various field strengths are sufficient
to determine the full set of polarisabilities.  As an example, the
behaviour of the correlator in the field $A^\mu_{(1)}(x)=(i a_1
\tau,0,0,0)$ (which corresponds to a constant electric field in the
$x_1$ direction) is given by
\begin{eqnarray}
  C_{s s^\prime}(\vec{p},\tau;A_{(1)}) &\sim&
  \delta_{s,s^\prime}\exp\left\{
- \frac{a_1\,\tau}{6M}\left[ a_1
  \left(q^2 \tau^2 + 12 M \pi  \alpha \right) - 3 i q\, \tau\,p_1\right]\right\}
e^{-M\,\tau} e^{-\frac{\tau}{2M}|\vec{p}|^2} +{\cal O}(a_1^3)
\nonumber
\\
&\stackrel{|\vec{p}|\to0}{\longrightarrow}&
\delta_{s,s^\prime}\exp\left[-(M+2\pi\alpha
  a_1^2)\tau-\frac{q^2a_1^2}{6M}\tau^3\right] +{\cal O}(a_1^3) \,.
  \label{eq:A1field}
\end{eqnarray}
In this case, the perturbative series has been resummed exactly in the
continuum, infinite volume limit and the higher order corrections come
from terms omitted in Eq.~(\ref{eq:eff_L}). For electrically neutral
particles, the exponential fall-off of this correlator determines the
polarisability $\alpha$ once the mass $M$ has been measured in the
zero-field case.  When a charged particle is placed in such a field it
undergoes continuous acceleration in the $x_1$ direction (this is
described by the $\tau^3$ term in the exponent).  However at times
small compared to $\sqrt{6}M/q\, a_1$, the correlator
essentially falls off exponentially. Matching the behaviour of
Eq.~(\ref{eq:A1field}) to lattice data for a charged hadron will again
enable us to determine the electric polarisability, $\alpha$. Whilst
the above analysis assumed infinite extent in the $x_1$ direction, it
remains valid for $L\ll m_\pi^{-1}$.

\section{Finite volume effects in nucleon polarisabilities}
\label{sec:hbxpt}

To calculate the quark mass and volume dependence of the nucleon
polarisabilities, we use heavy baryon chiral perturbation theory. Our
results also include the quenched, and partially-quenched versions of
this theory appropriate for existing lattice calculations.  The
details of the calculations are given in Ref.~\cite{Detmold:2006vu},
here we merely summarise the result for the case of the fourth spin
polarisability in the QCD case. It is convenient to separate the
different contributions as
\begin{eqnarray}
  \gamma_4= \gamma_4^{\rm anomaly} + \gamma_4^\Delta +\gamma_4^{\rm loop} \,,
\end{eqnarray}
corresponding to the contributions from the anomalous decay of flavour
neutral mesons to two photons, Born-terms involving the
$\Delta$-isobar resonance, and loop diagrams, respectively.

The anomalous decay of flavour neutral mesons to two photons has
important consequences in Compton scattering in non-forward
directions.  Anomalous decays are well understood in \CPT, entering
through the Wess-Zumino-Witten Lagrangian \cite{WZW}; extensions to
the partially-quenched case are discussed in
Ref.~\cite{Detmold:2006vu}.  The contributions to the amplitude from
the Born-terms involving the $\Delta$-isobar resonance are also
significant.  The resultant terms are found to be
\begin{eqnarray}
        \gamma_4^{\rm anomaly} = -\frac{3e^2 G_{\rm anom}}{16\pi^3
                f^2m_\pi^2} \,,\quad\quad
\gamma_4^\Delta =  \frac{e^2\mu_T^2}{216 \pi ( 2 M_N)^2 \Delta^2}\,,
\end{eqnarray}
where $G_{\rm anom}= g_A(2Z-1)$ and$\mu_T$ is the magnetic dipole
transition coupling \cite{Detmold:2006vu}.

For the loop contribution, using the effective couplings
$G_B=\frac{4}{3}g_A^2$ and $G_T=\frac{4}{9}g_{\Delta N}^2$ (which are
modified in the quenched and partially-quenched theories
\cite{Detmold:2006vu}), we find that \cite{Hemmert:1997tj}:
\begin{eqnarray}
 \gamma_4^{\rm loop}&=&
-\frac{e^2}{192\pi^3 f^2}\left[\frac{ G_B}{4m_\pi^2} 
+\frac{G_T}{3} \left\{\frac{1}{\Delta^2 - m^2} + \frac{\Delta}{2
    (\Delta^2 - m^2)^{3/2}} \ln 
\left[\frac{\Delta -
      \sqrt{\Delta^2 - m^2 + i \epsilon}}{\Delta + 
  \sqrt{\Delta^2 - m^2 + i \epsilon}}\right]\right\}
\right].\hspace*{3mm}
\end{eqnarray}

The finite volume of a lattice simulation restricts the available
momentum modes and consequently the results differ from their infinite
volume values.  Here we shall consider a hyper-cubic box of dimensions
$L^3\times T$ with $T\gg L$ and $m_\pi L\gg 1$ (smaller volumes in
which $m_\pi L\sim1$ \cite{Gasser:1987ah,Detmold:2004ap} are also
discussed in Ref.~\cite{Detmold:2006vu}) which leads to quantised
momenta $k=(k_0,{\vec k})$, ${\vec k}=\frac{2\pi}{L} {\vec
  j}=\frac{2\pi}{L} (j_1,j_2,j_3)$ with $j_i\in \mathbb{Z}$, but $k_0$
treated as continuous.  On such a finite volume, spatial momentum
integrals are replaced by sums over the available momentum modes.
Repeating the calculation of the loop diagrams using finite volume
sums rather than integrals leads to the following expression for the
loop contributions to $\gamma_4$:
\begin{eqnarray}
  \label{eq:gamma3_fv}
  \gamma_4^{\rm loop}(L) &=& - \frac{7e^2}{1152 \pi f^2}\int_0^\infty
d\lambda\Big[
3G_B {\cal F}_{\gamma_4}({\cal M})
-4 G_T {\cal F}_{\gamma_4}({\cal M}^\Delta)
\Big]\,,
\end{eqnarray}
where ${\cal M}=\sqrt{m_{\pi}^2+\lambda^2}$ and ${\cal
  M}^\Delta=\sqrt{m_{\pi}^2+2\lambda \Delta +\lambda^2}$ and the
finite volume sums ${\cal F}_\alpha(m) $ and ${\cal F}_{\gamma_4}(m)$
are defined in Ref.~\cite{Detmold:2006vu}.

\begin{figure}[!t]
\centering
\vspace*{-1cm}
        \includegraphics[width=0.7\columnwidth]{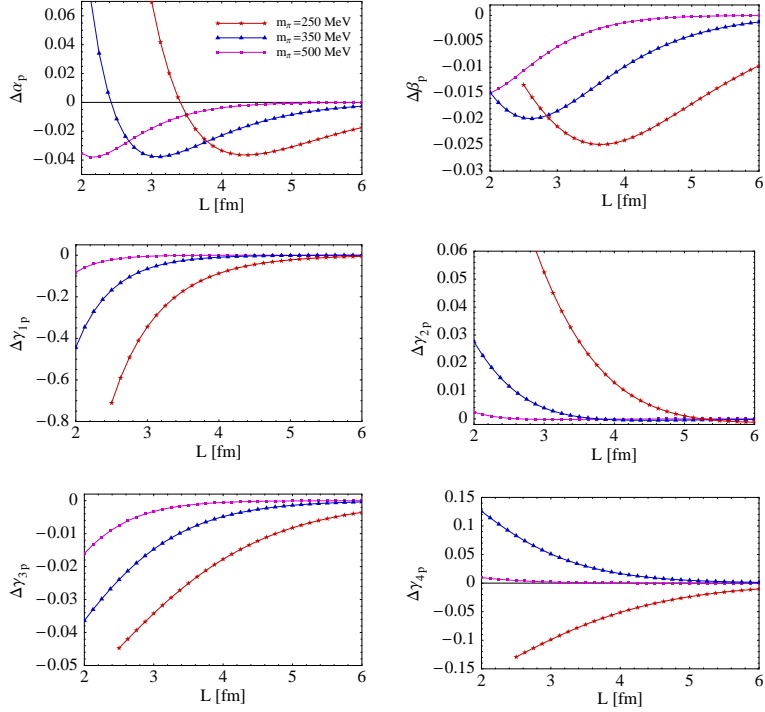}
\vspace*{-1cm}
\caption{Volume dependence of the proton polarisabilities. Here we
  show the ratio of the difference of the finite and infinite volume
  results to the infinite volume results for three values of the pion
  mass using the parameters described in the text. The curves
  terminate at $m_\pi\ L=3$.}
\label{fig:FV1}
\end{figure}

To illustrate the effects of the finite lattice extent, Figure
\ref{fig:FV1} shows the volume
dependence of the various polarisabilities in the proton. Here we have
specialised to QCD, and chosen physical values for the various
constants.  In the figure, we show results for the ratio $\Delta
X(L)=[X(L)-X(\infty)]/X(\infty)$ for the six polarisabilities at three
different pion masses, $m_\pi=0.25,\,0.35,\,0.50$~GeV. The overall
magnitude of these shifts varies considerably; generally volume
effects are at the level of 5--10\% for $m_\pi=0.25$~GeV and smaller
for larger masses. Larger effects are seen in a number of the spin
polarisabilities but there are as yet no lattice calculations of these
quantities.  Results in Ref.~\cite{Detmold:2006vu} also allow us to
calculate the finite volume effects in the quenched data on the
various polarisabilities calculated in
Refs.~\cite{Christensen:2004ca,Lee:2005dq}. The quenched expressions
involve a number of undetermined LECs, so we can only estimate the
volume effects. Using reasonable range for these parameters, we see
that the calculations on a (2.4 fm)$^3$ lattice with $m_\pi=$0.5~GeV
may differ from their infinite volume values by 5--10\% in the case of
the electric polarisability.

\section{Conclusion}
\label{sec:discussion}

We have investigated Compton scattering from spin-half targets from
the point of view of lattice QCD. We first discussed how external
field methods can be used to probe all six polarisabilities of real
Compton scattering for both charged and uncharged targets. We also
discussed the effects of the finite volume used in lattice
calculations on the polarisabilities. Since polarisabilities are
infrared-sensitive observables (they scale as inverse powers of the
pion mass near the chiral limit), they are expected to have strong
volume dependence. This is indeed borne out in the explicit
calculations presented here. In QCD, we generically find that the
polarisabilities experience volume shifts of 5--10\% from the infinite
volume values for lattice volumes $\sim$(2.4~fm)$^3$ and pions of mass
0.25~GeV. In the case of quenched QCD (relevant to the
only existing lattice data), we find significant effects even at pion
masses $\sim0.5$~GeV. Future lattice studies of the polarisabilities
should take these effects into account in order to present physically
relevant results.

\acknowledgments This work is supported by the US DOE: 
DE-FG02-97ER41014 and DE-FG02-05ER41368-0.


\begin{thebibliography}{99}

\bibitem{Detmold:2006vu}
  W.~Detmold, B.~C.~Tiburzi and A.~Walker-Loud,
  %``Electromagnetic and spin polarisabilities in lattice QCD,''
  Phys.\ Rev.\ D {\bf 73}, 114505 (2006).
  %%CITATION = HEP-LAT 0603026;%%


%\cite{Ragusa:1993rm}
\bibitem{Ragusa:1993rm}
  S.~Ragusa,
  %``Third Order Spin Polarizabilities Of The Nucleon,''
  Phys.\ Rev.\ D {\bf 47}, 3757 (1993).
  %%CITATION = PHRVA,D47,3757;%%


%\cite{Fiebig:1988en}
\bibitem{Fiebig:1988en}
  H.~R.~Fiebig, W.~Wilcox and R.~M.~Woloshyn,
  %``A STUDY OF HADRON ELECTRIC POLARIZABILITY IN QUENCHED LATTICE QCD,''
  Nucl.\ Phys.\ B {\bf 324}, 47 (1989).
  %%CITATION = NUPHA,B324,47;%%

%\cite{Christensen:2004ca}
\bibitem{Christensen:2004ca}
  J.~Christensen, W.~Wilcox, F.~X.~Lee and L.~m.~Zhou,
  %``Electric polarizability of neutral hadrons from lattice QCD,''
  Phys.\ Rev.\ D {\bf 72}, 034503 (2005).
  %%CITATION = HEP-LAT 0408024;%%

%\cite{Lee:2005dq}
\bibitem{Lee:2005dq}
  F.~X.~Lee, L.~Zhou, W.~Wilcox and J.~Christensen,
%   ``Magnetic polarizability of hadrons from lattice QCD in the background
  %field method,''
  Phys.\ Rev.\ D {\bf 73}, 034503 (2006).
  %%CITATION = HEP-LAT 0509065;%%

%\cite{Weinberg:1966kf}
\bibitem{CHPT}%\bibitem{Weinberg:1966kf}
  S.~Weinberg,
  %``Pion scattering lengths,''
  Phys.\ Rev.\ Lett.\  {\bf 17}, 616 (1966);
  %%CITATION = PRLTA,17,616;%%
%
%\cite{Gasser:1983yg}
%\bibitem{Gasser:1983yg}
  J.~Gasser and H.~Leutwyler,
  %``Chiral Perturbation Theory To One Loop,''
  Annals Phys.\  {\bf 158}, 142 (1984);
  %%CITATION = APNYA,158,142;%%
%
%\cite{Gasser:1984gg}
%\bibitem{Gasser:1984gg}
%  J.~Gasser and H.~Leutwyler,
  %``Chiral Perturbation Theory: Expansions In The Mass Of The Strange Quark,''
  Nucl.\ Phys.\ B {\bf 250}, 465 (1985).
  %%CITATION = NUPHA,B250,465;%%


%\cite{Fucito:1982ff}
\bibitem{Fucito:1982ff}
  F.~Fucito, G.~Parisi and S.~Petrarca,
%   ``First Evaluation Of G(A) / G(V) In Lattice QCD In The Quenched
  %Approximation,''
  Phys.\ Lett.\ B {\bf 115}, 148 (1982).
  %%CITATION = PHLTA,B115,148;%%

%\cite{Martinelli:1982cb}
\bibitem{Martinelli:1982cb}
  G.~Martinelli, G.~Parisi, R.~Petronzio and F.~Rapuano,
  %``The Proton And Neutron Magnetic Moments In Lattice QCD,''
  Phys.\ Lett.\ B {\bf 116}, 434 (1982).
  %%CITATION = PHLTA,B116,434;%%

%\cite{Bernard:1982yu}
\bibitem{Bernard:1982yu}
  C.~W.~Bernard, T.~Draper, K.~Olynyk and M.~Rushton,
  %``Lattice QCD Calculation Of Some Baryon Magnetic Moments,''
  Phys.\ Rev.\ Lett.\  {\bf 49}, 1076 (1982).
  %%CITATION = PRLTA,49,1076;%%

%\cite{Aoki:1989rx}
\bibitem{EDM}%\bibitem{Aoki:1989rx}
  S.~Aoki and A.~Gocksch,
  %``The Neutron Electric Dipole Moment In Lattice QCD,''
  Phys.\ Rev.\ Lett.\  {\bf 63}, 1125 (1989)
  [Erratum-ibid.\  {\bf 65}, 1172 (1990)];
  %%CITATION = PRLTA,63,1125;%%
%
%\cite{Aoki:1990ix}
%\bibitem{Aoki:1990ix}
  S.~Aoki, A.~Gocksch, A.~V.~Manohar and S.~R.~Sharpe,
  %``Calculating the neutron electric dipole moment on the lattice,''
  Phys.\ Rev.\ Lett.\  {\bf 65}, 1092 (1990).
  %%CITATION = PRLTA,65,1092;%%

%\cite{Shintani:2005du}
\bibitem{Shintani:2005du}
  E.~Shintani {\it et al.},
  %``Neutron electric dipole moment on the lattice,''
  PoS {\bf LAT2005}, 128 (2006).
  %%CITATION = HEP-LAT 0509123;%%

%\cite{Detmold:2004kw}
\bibitem{Detmold:2004kw}
  W.~Detmold,
  %``Flavour singlet physics in lattice QCD with background fields,''
  Phys.\ Rev.\ D {\bf 71}, 054506 (2005).
  %%CITATION = HEP-LAT 0410011;%%

%\cite{Babusci:1998ww}
\bibitem{Babusci:1998ww}
  D.~Babusci, G.~Giordano, A.~I.~L'vov, G.~Matone and A.~M.~Nathan,
%   ``Low-energy Compton scattering of polarized photons on polarized
  %nucleons,''
  Phys.\ Rev.\ C {\bf 58}, 1013 (1998).
  %%CITATION = HEP-PH 9803347;%%

%\cite{Schwinger:1951nm}
\bibitem{Schwinger:1951nm}
  J.~S.~Schwinger,
  %``On gauge invariance and vacuum polarization,''
  Phys.\ Rev.\  {\bf 82}, 664 (1951).
  %%CITATION = PHRVA,82,664;%%


%\cite{Wess:1971yu}
\bibitem{WZW}%\bibitem{Wess:1971yu}
  J.~Wess and B.~Zumino,
  %``Consequences of anomalous Ward identities,''
  Phys.\ Lett.\ B {\bf 37}, 95 (1971);
  %%CITATION = PHLTA,B37,95;%%
%
%\cite{Witten:1983tw}
\bibitem{Witten:1983tw}
  E.~Witten,
  %``Global Aspects Of Current Algebra,''
  Nucl.\ Phys.\ B {\bf 223}, 422 (1983).
  %%CITATION = NUPHA,B223,422;%%

%\cite{Hemmert:1997tj}
\bibitem{Hemmert:1997tj}
  T.~R.~Hemmert, B.~R.~Holstein, J.~Kambor and G.~Knochlein,
%   ``Compton scattering and the spin structure of the nucleon at low
  %energies,''
  Phys.\ Rev.\ D {\bf 57}, 5746 (1998).
  %%CITATION = NUCL-TH 9709063;%%

%\cite{Gasser:1987ah}
\bibitem{Gasser:1987ah}
  J.~Gasser and H.~Leutwyler,
  %``THERMODYNAMICS OF CHIRAL SYMMETRY,''
  Phys.\ Lett.\ B {\bf 188}, 477 (1987).
  %%CITATION = PHLTA,B188,477;%%

%\cite{Detmold:2004ap}
\bibitem{Detmold:2004ap}
  W.~Detmold and M.~J.~Savage,
  %``Nucleon properties at finite volume: The epsilon'-regime,''
  Phys.\ Lett.\ B {\bf 599}, 32 (2004).
  %%CITATION = HEP-LAT 0407008;%%

\end{thebibliography}
\end{document}